\documentclass[12pt,preprint]{aastex}

\begin{document}

\shorttitle{The flattened dark matter halo of M31}
\shortauthors{Banerjee \& Jog}
\title{The flattened dark matter halo of M31 as deduced from the observed HI scale heights}

\author{Arunima Banerjee \&  Chanda J. Jog}
\affil{Department of Physics, Indian Institute of Science, Bangalore 560012, India}
\email{arunima$\_$banerjee@physics.iisc.ernet.in, cjjog@physics.iisc.ernet.in}

\begin{abstract}
In this paper, we use the outer-galactic HI scale height data as well as the observed
rotation curve as constraints to determine the halo density distribution of the Andromeda galaxy (M31). We model the galaxy as a gravitationally-coupled system of stars and gas, responding to the external force-field of a known Hernquist bulge and the dark matter halo, the density profile of the latter being characterized by four free parameters. The parameter space of the halo is optimized so as to match the observed HI thickness distribution as well as the rotation curve on an equal footing, unlike the previous studies of M31 which were based on rotation curves alone. We show that an oblate halo, with an isothermal density profile, provides the best fit to the observed data. This gives a central density of 0.011 M$_{\odot}$ pc$^{-3}$, a core radius of 21 kpc, and an axis ratio of 0.4. The main result from this work is the flattened dark matter halo for M31, which is required to match the outer galactic HI scale height data. 
 Interestingly, such 
flattened halos lie at the most oblate end of the distribution of halo shapes found in recent cosmological simulations.\\
\end{abstract}
\keywords{galaxies: ISM - galaxies:
kinematics and dynamics -  galaxies: spiral -
galaxies: structure - galaxies: halos}

\section{Introduction}
It is well-known that the dark matter halo plays an important role in the dynamics of galaxies, especially in the outer regions (Binney \& Tremaine 1987). Since a galactic disk is rotationally supported, the  rotation curve serves as a useful tracer of the gravitational potential in the plane of the galaxy. The observed rotation curve is routinely used to deduce the mass distribution in a galaxy and hence its dark matter content (e.g., Begeman 1987, Kent 1986, 1987, Geehan et al. 2006). The thickness of the gas layer, on the other hand, depends on the vertical gravitational force and traces the potential perpendicular to the mid-plane (e.g., Narayan \& Jog 2002 a). 
In this work, we use the rotation curves as well as the radial distribution of the thickness of the HI gas layer in the outer galaxy to study the shape and density profile of the dark matter halo in M31. 
In  a disk plus bulge plus halo model of an external  galaxy, the disk and the bulge can be mostly studied observationally. Therefore, the rotation curve and the vertical
HI scale height
 data effectively complement each other to determine the dark matter halo distribution of a galaxy uniquely.

 In the past, the idea of studying the dark matter halo properties by using the outer galactic HI flaring data has been used to explore tha halos of NGC 4244 (Olling 1996), NGC 891 (Becquaert \& Combes 1997), and the Galaxy (Olling \& Merrifield 2000, 2001). However, the HI scale height distribution was mainly used to constrain the oblateness of the halo, and not its other parameters such as the power-law index. In some cases, the gas gravity and even the stellar gravity was ignored (Becquaert \& Combes 1997) in determining the net galactic potential, and hence the gas scale height distribution.

These issues were taken care of in determining the Galactic halo parameters by Narayan et al (2005). Using the gravitationally-coupled, 3-component Galactic disk model (Narayan \& Jog 2002 b), various density profiles of the halo were investigated, and an attempt was made to obtain the halo parameters, which provided the best fit (in the least square sense) to the observed HI scale height distribution. Finally, conformity with the shape of the observed rotation curves was used to remove the degeneracies in the best-fit values obtained by the first constraint. Also, unlike some of the previous models, the self-gravity of the gas was included in the analysis. From their study Narayan et al. (2005) concluded that a spherical halo, with a density falling off more rapidly than an isothermal halo,  provides the best fit to the available data. This study was based on the HI scale height data then available upto 24 kpc from Wouterloot et al. (1990). Kalberla et al. (2007) have confirmed this by using their recent extended HI scale
height data upto 40 kpc, and have also included a dark matter ring which they claim is needed to explain the observed HI scale height distribution in the Galaxy.

In this paper, we apply the above approach
 to investigate the dark matter halo properties of the Andromeda galaxy (M31 or NGC 224). Here we use both the rotation curve and the HI scale height data as rigorous constraints simultaneously
and scan the entire parameter space
 systematically so as to obtain the best-fit halo parameters. In addition to the various density profiles, we also
 try to fit various shapes of the halo which was not done by Narayan et al. (2005).
Earlier studies on M31  (Widrow et al. 2003, Widrow \& Dubinski 2005, Geehan et al. 2006, Seigar et al. 2006, Tamm et al. 2007) were mostly aimed at developing a complete mass model (disk plus bulge plus halo), based on comparisons made with the available structural and kinematical data (surface brightness profiles, bulge-velocity dispersion relations, rotation curves), which assumed a spherical-shaped halo. On the other hand, we have studied the 
dark matter halo profile, and  show it to be flattened.

In Section 2, we describe
 our model, and in  Section 3 discuss the numerical calculations involved, and the input parameters used. In Section 4, we present the results and analysis of the numerical results; followed by the discussion,
and conclusions in Sections 5 and 6 respectively.

\section{Details of the model}

\noindent {\bf 1. Gravitationally coupled, 3-component, galactic disk model}

A galaxy is modelled as a thick stellar disk, coplanar with the interstellar medium of atomic and molecular hydrogen gas, all the three components being gravitationally coupled to each other, and embedded in the dark matter halo (Narayan \& Jog 2002 b). The bulge and the dark matter halo are taken to be rigid and non-responsive, and act as external forces on the 3-component disk system. Also, it is assumed that the components are in hydrostatic equilibrium in the vertical direction. Therefore, the density distribution of each component will be jointly determined by the Poisson equation, and the corresponding equation for pressure equilibrium perpendicular to the midplane.

The Poisson equation for an axisymmetric  galactic system in terms of the galactic cylindrical co-ordinates (R, $\phi$, z) is given by \\
$$\frac{{\partial}^2{\Phi}_{total}}{{\partial}z^2} + \frac{1}{R}\frac{{\partial}}{{\partial}R}(R \frac{{\partial}\Phi_{total}}{{\partial}R})
 = 4\pi G(\sum_{i=1}^{3} \rho_i+ \rho_{b} + \rho_{h})
\eqno(1) $$ \\
where $\rho_i$ with i = 1 to 3 denotes the mass density for each disk component. $\rho_h$ and $\rho_b$ denote the same for the halo and the bulge respectively. $\Phi_{total}$ denotes the net potential due to the disk, halo and the bulge. For a flat or gently-falling rotation curve, the radial term can be neglected as its contribution to the determination of the HI scale height is less than 10 percent as noted by earlier calculations (Narayan et al. 2005). So, the above equation reduces to \\
$$\frac{{\partial}^2\Phi_{total}}{{\partial}z^2}
 = 4\pi G(\sum_{i=1}^{3} \rho_i + \rho_{b} + \rho_{h})
\eqno(2) $$ 
The equation for hydrostatic equilibrium in the z direction is given by (Rohlfs 1977)
$$ \frac{\partial}{{\partial}z}(\rho_{i}<(v_{z}^{2})_{i}>) + \rho_{i}\frac{{\partial}\Phi_{total}}{{\partial}z} = 0  \eqno(3) $$ \\
\noindent where $<(v_{z}^{2})_{i}>$ is the mean square random velocity along the $z$ direction for the component $i$. We further assume each component to be isothermal for simplicity, so that the velocity term is constant with $z$. 

Eliminating $\Phi_{total}$ between eq. (2) and eq. (3), and assuming an isothermal case, we 
get
$$ <(v_{z}^{2})_{i}>  \frac{\partial}{{\partial}z}[\frac{1}{\rho_{i}}\frac{{\partial}\rho_{i}}{{\partial}z}] = -4\pi G(\sum_{i=1}^{3} \rho_i+ \rho_{b} + \rho_{h})
 \eqno(4)$$
which represents a set of three coupled, second-order differential equations, one for each component of the disk. From the above equation, it is evident that though there is a
common gravitational potential, the response of each component
will be different due to the difference in their random velocity
dispersions.

\noindent {\bf 2. Bulge}

We model the bulge of M31 as a spherically symmetric mass distribution represented by a Hernquist profile (Hernquist 1990),
where $M_b$ is the total mass of the bulge, and $r_b$ is its core radius. The mass profile and density corresponding to this distribution are given by \\

$$ M_{bulge}(R) = \frac{M_b  R^2}{(r_b + R)^2} \eqno(5) $$ \\
and
$$ {\rho}_{bulge}(R) = \big( \frac{M_b}{2\pi r_b^3} \big) \frac{1}{(R/r_b)(1+R/r_b)^3} \eqno(6) $$ \\
respectively. Since M31 is an Sb type galaxy, the bulge contribution is important, and plays a role in determining the rotation curve even in regions outside the bulge.

\noindent {\bf 3. Dark Matter Halo} 

We use the four-parameter dark matter halo model (de Zeeuw \& Pfenniger 1988; Becquaert
\& Combes 1997) with the density profile given by

$$\rho(R,z) = \frac{\rho_0}{\large [ 1+\frac{m^{2}}{{{R_c}(q)}^{2}}\large]^p} \eqno(7) $$  \\
where $ m^{2}$=$R^{2} + ({z^{2}}/{q^{2}})$, 
 $\rho_0$  is the central core density of the halo, ${R_c}(q)$
is the core radius, $p$ is the density index, and $q$ is the vertical-to-planar axis ratio of the halo 
(spherical: $q$ = 1; oblate: $q$ $<$ 1; prolate: $q$ $>$ 1).

\section {Numerical Calculations}

\subsection{Solution of equations}
For a given bulge and halo density profile, the equation to be solved to obtain the vertical density distribution at any radius for any component (stars, HI \& $H_2$) is given by eq.(4), which simplifies to: 
$$ \frac{d^2\rho_i}{dz^2} = \frac{\rho_i}{<({v_z}^2)_i>} \times [ -4\pi G(\rho_s+\rho_{HI}+\rho_{H_2} + \rho_{b}+\rho_{h}) ] +\frac{1}{\rho_i} (\frac{d\rho_i}{dz})^2  \eqno(8) $$ \\

This represents three coupled, second-order, ordinary differential equations in $\rho_s$, $\rho_{HI}$ and $\rho_{H_2}$
which denote the mass densities for stars, HI and H$_2$ respectively. This problem is solved in an iterative fashion, as an initial value problem, using fourth order, Runge-Kutta method of integration, with the following two initial conditions at the
mid-plane i.e z = 0 for each component:
$$ \rho_i = (\rho_0)_i,  \qquad \frac{d\rho_i}{dz} = 0  \eqno(9) $$ \\
However, the modified mid-plane density $(\rho_0)_i $ for each
component is not known a priori. Instead the net surface
density $\Sigma_i(R)$,
given by twice the area under curve of
$\rho_i(z)$ versus z, is used as the second boundary condition,
since this is known observationally.
Hence the required value of $(\rho_i)_0$
can be determined by trial and error method, which eventually
fixes the $\rho_i(z)$ distribution. We find that four iterations are adequate to
give convergence with an accuracy to the second decimal point. For a three-component disk, the
vertical distribution is steeper than a sech$^2$ close to the mid-plane (Banerjee \& Jog 2007) but at large $z$ values it is close to a sech$^2$ distribution. Here we use the HWHM (half width at half maximum) of the resulting model vertical distribution to define the scale height as was done in Narayan \& Jog (2002 a, b).

\subsection{Input Parameters for M31}

The model described so far is general, now we apply it to M31. For that, we
require the surface density and the vertical velocity dispersion for each component to solve the coupled set of equations at each radius. The observed values are used for the gas, whereas for the bulge and the stellar component the values derived from the mass-model are used except for the vertical dispersion velocity of the stars. 

The radial distribution of the HI surface density is taken from Sofue \& Kato (1981), while the surface density for the $H_2$ gas, being an order of magnitude smaller and confined to the inner region (Koper 1993), was taken to be zero. The vertical dispersion velocity of the HI gas,
$(v_z)_{HI}$, is taken to be 8 km s$^{-1}$, as given by the observations of a large sample of about 200 external galaxies (Lewis 1984). 

For the stars, we assume an exponential disk with a central surface density of 460 M$_{\odot}$ pc$^{-2}$, and an exponential disk scale-length, R$_d$ of 5.4 kpc after Geehan et al. (2006). The disc scale-length matches with the R-band scale-length of Widrow et al. (2003), whereas Koper (1993) gives a value of 5.1 kpc. Also, the maximum disk mass of 8 $\times$ $10^{10}$ M$_{\odot}$ predicted by the Widrow et al. (2003) model sets the central disk surface density value to 440 M$_{\odot}$ pc$^{-2}$ for $R_{d}$ = 5.4 kpc. So the difference between the values used in our work with other values in the literature is of the order of a few percent.

The bulge is taken to have a Hernquist profile as described earlier,
with a total mass $M_b$ of 3.3 $\times$ $10^{10}$ M$_{\odot}$ and core radius $r_b$ of 0.61 kpc, as given by the same model (Geehan et al. 2006).
 Kerins et al. (2001) give a mass
$M_b$ of 4 $\times$ $10^{10}$ M$_{\odot}$, whereas Widrow et al. (2003) give a mass 
of 2.5 $\times$ $10^{10}$ M$_{\odot}$. As expected, the bulge becomes progressively less
important as we move radially outward in the galaxy, and so is not important in the determination of the rotation curve or the HI scale height in the outer parts,
which is our region of interest. However, it does affect the determination of the rotation curve in the intermediate range, hence the bulge has to be included for a correct treatment of the problem.

The stellar radial velocity dispersion is assumed to fall off exponentially with a scale-length of  2  $\times$ $R_{d}$  as is observed in the Galaxy (Lewis \& Freeman 1989), for M31 this gives a value of 10.8 kpc. Also, the ratio of the vertical to the radial stellar velocity dispersion is taken to be 0.5 at all radii, equal to its observed value in the solar neighbourhood
(Binney \& Merrifield 1998). Based on these, the
central value of 126 km s$^{-1}$ for the radial dispersion is deduced 
from the observed value at 2 R$_d$ for M31 by Tamm et al. (2007).

\section{Results \& analysis}

\subsection{Halo density profiles: The 3-D Grid}

We scan the allowed values for the entire parameter space for the dark matter halo, to obtain the best fits to 
the data for the rotation curve and the HI scale-heights in the outer parts of M31.

We vary the halo density index $p$ between 1, 1.5 and 2. Here p = 1 corresponds to the isothermal and $p$ = 1.5 to the NFW (Navarro et al. 1996) halo density profile at large radii. These two profiles are routinely used in galactic mass modelling and other studies.  Narayan et al.(2005) found that a steeper-than NFW profile ($p$ = 2) best conforms with the  HI scale height data in the outer regions of our Galaxy, suggesting the evidence of finite-sized halos. So, we study each of the above three cases individually. For each value of $p$, a realistic range of $\rho_0$ and $R_c$ for the spherical case is chosen to form a grid of \{$\rho_0$, $R_c$\} pairs. $\rho_0$  is varied between 0.001 - 0.15 M$_{\odot}$ pc$^{-3}$ in steps of 0.002 M$_{\odot}$ pc$^{-3}$, and $R_c$ is varied between 1 - 35 kpc in steps of 0.5 kpc. At first, we use a spherical halo for simplicity. Although it gives the correct rotation curve, it fails to match the outer galactic HI scale-height data within the error bars. Figure 1  shows that the outer galactic HI scale height distribution obtained by using the best-fit values for a spherical isothermal halo ($p$=1) flares far above the observed data, giving a poor fit. The same is found to be true for the other density profiles ($p$ = 1.5 and 2) as well.

This naturally called for the use of an oblate (flattened) halo. This is because the midplane surface density increases with the flattening of the halo, thus resulting in a higher vertical constraining force. The effect, as expected, is exactly the opposite in case of a prolate halo. 
The latter would therefore give an even poorer fit than the spherical case, and hence is not tried here.
Therefore, in addition to the above parameters, the axis ratio $q$ is varied as well between 0.1 and 0.9 in steps of 0.1. This gives a total of 47250 grid points to be scanned for each value of $p$. We first thoroughly scan this grid to locate the region of minimum ${\chi}^2$. 

In retrospect, Narayan et al. (2005) had pinned the rotation curve at a single point only (i.e the solar point with R = 8.5 kpc) using the local Oort constants $A$ and $B$, for which the values are available for the Galaxy. This effectively fixed the rotation curve locally, with respect to shape as well. Also, the global trends exhibited by the observed curve was used as the final criterion to choose the best fit density index ($p$ = 2). Here, on the other hand, we apply  a more rigorous treatment by pinning the rotation curve at all the observed points. This, in fact, was imperative since the Oort $A$ and $B$ constants for M31 are not known.

\subsection{The rotation curve constraint}

For each of the above grid points, we evaluate the galactic rotation curve using our (disc plus bulge plus halo) model as follows. For an exponential disk, the rotation velocity $v_{disk}(R)$ is given by (Binney \& Tremaine 1987)\\
$$ v_{disk}^{2}(R) = 4\pi G \Sigma_{0} R_{d} y^{2} [I_0(y)K_0(y) - I_1(y)K_1(y)] \eqno(10)$$ 	\\
where $\Sigma_{0}$ is the disk central surface density, $R_{d}$ the disc scale length and y = {R}/{2R$_{d}$}, R being the galactocentric radius. $I_{n}$ and $K_{n}$ (where n=0 and 1) are the modified Bessel functions of the first and second kind respectively.
The above relation is for an infinitesimally thin disk which we use here for simplicity.
For a thick disk, a separate result has to be used (as given in Becquaert \& Combes 1997) , which we check gives a value within $< 1 \% $ of the value given by eq.(10), hence we are justified in using the above simpler form.

For the spherical bulge, rotation velocity $v_{bulge}(R)$ is given by \\
$$ v_{bulge}^{2}(R) = \frac{G M_{bulge} (R)}{R} \eqno(11) $$ \\
where  M$_{bulge} (R)$, the mass enclosed within a sphere of radius R for a Hernquist bulge, is given by 
the r.h.s. of equation (5).

 For an oblate halo of axis ratio $q$ and density index $p$, the circular speed $v_{halo}$ is obtained by differentiating the expression for the potential from Sackett \& Sparke (1990), and Becquaert \& Combes (1997) to be: \\
$$ v_{halo}^{2}(R) = 4 \pi G \rho_{0} q \int_{0}^{1/q} \frac{R^2 x^2 [ 1 + \frac{R^2 x^2}{R_{c}^2 ( 1 + \epsilon^2 x^2)}  ]^{-p}}{( 1 + \epsilon^2 x^2)^2} dx						\eqno(12)$$ \\
\noindent where $\epsilon = (1- q^2)^{1/2}$.
We obtain the value of the integral numerically in each case.
So, the total rotation velocity $v(R)$ at any galactic radius $R$ is given by adding 
the contributions for the three components in quadrature as:\\
$$ {v^{2}(R)} = v_{disk}^{2}(R) + v_{bulge}^{2}(R) + v_{halo}^{2}(R) \eqno(13) $$ 		\\
Next we performed the ${\chi}^2$ analysis of the calculated distributions with respect to the observed one for each of the $p$ = 1, 1.5 and 2 cases. The galactic rotation curve for M31 is given by Widrow et al. (2003), and Carignan et al. (2006). We choose the rotation curve from Widrow et al. (2003) for our analysis following Geehan et al (2006) for reasons of internal consistency as we have used the bulge and the stellar parameters from their mass model. 
We find that the calculated rotation curves do not depend significantly on the shape of the halo, i.e.,  $q$. Also, all the three density indices ($p$ = 1, 1.5 and 2) give reasonably good fit to the observed rotation curve, in that they lie well within the error bars of the observed curve in the radial range of 2-30 kpc. However, in terms of ${\chi}^2$ analysis, p = 1 (the isothermal case) is the most favoured.

\subsection{The HI scale height constraint}

Now, for each of the above three cases ($p$ = 1, 1.5, 2), we considered only those grid points for HI scale height determination, whose ${\chi}^2$ values were less than 11, which is the total number of data points in the observed rotation curve.
Elementary statistics suggests that a model distribution can be taken to be a reasonably good fit if its ${\chi}^2$ value with respect to the observed distribution is of the order of the number of data points in the observed distribution (Bevington 1969). Also, for these given set of grid points, we have checked that the rotation curve lies well within the error bars in the region of interest i.e the outer galactic region. For $p$ = 1, there were 3684 such grid points, 4910 for $p$ = 1.5 and 5799 for $p$ = 2. 

For the set of grid points chosen above, we calculated the HI scale        
height in the outer galactic region,  i.e,  beyond three disc scale lengths (R = 
16.2 kpc) following the analysis done for the Galaxy case (Narayan et al 2005), and obtained the ${\chi}^2$ with respect to the observed scale height data (Braun 1991). The dark matter halo is expected to be more important in the outer parts and hence the scale height values
 in this range are used as a constraint to obtain the halo parameters.

\subsection{Resulting Best-fit halo parameters}

We find that a flattened, isothermal halo ($p$=1, $q$ = 0.4) with a  small central density (0.011 M$_{\odot}$ pc$^{-3}$), and a core radius of 21 kpc provides the best fit to both the observed rotation curve and the HI scale height data. 
The other density profiles ($p$ = 1.5, and 2) give higher values for the ${\chi}^2$ minimum and hence are not favoured, however, the dependence on the power-law index $p$ is weak.
Interestingly these also give $q$ = 0.4 as the preferred shape and the same central density but with a higher core radius of 27 kpc and 34 kpc respectively.

In Fig.2, we compare our best fit rotation curve with the observed one in the outer regions although the fit was done over the entire radial range of 2-30 kpc (Section 4.2). Similarly, Fig.3 depicts the best-fit scale height distribution in the outer disk region.
In Fig. 4, we compare the resulting scale height distributions for the various values of $q$ = 0.2 - 1 in steps 0.2, obtained using the best-fit values for ${\rho}_0$
and $R_c$ in each case. It shows that although the ${\chi}^2$ analysis selects $q$ = 0.4 to be the best fit, 
 a range of $q$ values between 0.4 - 0.6 gives fits to the observed data within the error bars - but the departure of the calculated distribution from the real data points increases with increasing $q$.
This is further illustrated in Figure 5, where we plot the ${\chi}^2$ value for the best-fit case for each axis ratio vs. the axis ratio, $q$. 
The ${\chi}^2$ shows a clear minimum at $q$ =0.4 which is thus the axis ratio that best explains the observations.
In particular,  a flattened halo ($q$ = 0.4 ) is clearly distinguished from and preferred  over the
spherical  case ($q$ = 1).

Interestingly, such 
flattened halos lie at the most oblate end of the distribution of halo shapes  obtained in recent $\Lambda$ CDM cosmological simulations
(e.g., Bailin \& Steinmetz 2005, see their Fig. 2; Bett et al. (2007), see their Fig. 13)- the axis ratio $q$ used in the present paper is equal to the ratio of the semi-major axes $c/a$ in these papers.
Thus, either M31 is an unsual galaxy, or the simulations need to include additional physics such as the effect of baryons that could affect the shape of the halo. Further, a moderate variation in HI gas dispersion results in a less flattened halo as shown in Section 5, point 3.

\section{Discussion}

\noindent {\bf 1. HI Scale height data in outer disk:} The HI scalehight constraint as applied in this paper is ideally suited for application to gas-rich, late-type spiral galaxies with an extended HI disk. In order to be useful as a constraint, the HI scale height data should be available beyond 3 - 4 $R_{d}$ and even farther out in the galaxy. This is where the disk gravitational force begins to drop out and the dark matter halo takes over. 
We note that obtaining the HI scale height data is an observationally challenging task (Sancisi \& Allen 1979), and therein lies the main difficulty in using this method. As far as our work is concerned, the observational data (Braun 1991) gives only three data points beyond R = 3$R_{d}$. We consider the region only beyond R = 3$R_{d}$ following the Galaxy case (Narayan et al. 2005). Also, the irregularity or the scatter in the observed data in the inner region suggests the presence of a bar or spiral arm or some unknown structure, and is therefore excluded from the analysis.   

In Fig.6 we illustrate the above point by plotting the halo surface density within the HI scale height, as well as the 
corresponding values for the bulge, stars, and HI gas versus the radius. For using the HI scale height constraint, the vertical force close to the galactic mid-plane is needed, this is why the
surface density of the halo within the HI scale height is included. This figure shows that the halo surface density just begins to take over the stellar density at the point beyond which we do not have any observed data. Availability of more data points in the outer parts, with lower error-bars, is thus clearly desirable and would yield a tighter constraint on the halo shape and the density profile.

\noindent {\bf 2. Simultaneous constraints:} The calculated rotation curve does not depend on the shape of the halo ($q$). All the $q$ values 
give equally good fits to the observed data. Surprisingly, the ${\chi}^2$ minima for the rotation curve 
and the HI scale height data, taken separately, lie on different regions of the grid. The best-fit to the rotation curve alone gives high values of the central density and small values of core radius. The best-fit to the scale height data, on the other hand, gives a  lower central density and a larger core radius. Geehan et al. (2006) obtained a best-fit $\rho_{0}$ of 0.033 M$_{\odot}$ pc$^{-3}$ (somewhat higher than the value we get)
and $R_{c}$ of 8.2 kpc , probably because they had used the rotation curve as the only constraint. We, on the other hand, have used two complementary constraints; the planar one involving the rotation curve, and the vertical one involving the HI scale height. This allows us to uniquely determine the physical parameters of the halo, and the determination of the true minimum in the grid spanning all the parameters for the dark matter halo. 

We find that the observed HI scale height
increases linearly with radius and does not flare, and this can be explained if the
halo is flattened  and consequently implies a smaller central density, while keeping the same
mass within a radius to explain the observed rotation curve.

\noindent {\bf 3. Dependence on the HI velocity dispersion:} The calculation of the HI scale height is crucially dependent on the chosen value of the vertical velocity dispersion of HI. We do not have any observationally measured value for the gas dispersion for M31. Here we have assumed it to be constant at 8 km s$^{-1}$ at all radii, as is observed for about 200   galaxies (Lewis 1984). If at all, the observed dispersion falls off with radius very gradually in the outer galaxy (Kamphuis 1993, Narayan et al. 2005).
Repeating the whole analysis for a slightly smaller value of $(v_z)_{HI}$ = 7 km s$^{-1}$, we find the best fit $\rho_{0}$ and $R_{c}$ to be 0.007 M$_{\odot}$ pc$^{-3}$ and 19.5 kpc respectively with $q$ = 0.8-0.9, implying a less oblate halo, but with  larger ${\chi}^2$ values, than obtained for the best-fit case with 8 km s$^{-1}$ in Section 4.4.
Physically this trend can be explained as follows: in this case the pressure support is smaller hence the gravitational force required to balance it has to be smaller to match the same observed data, hence a less flattened halo is expected. 

We also tried a case with a small radial fall-off in the HI dispersion from 8 to 7 km s$^{-1}$ in the outer disk between 18-27 kpc. This variation assumed is ad-hoc, but it allows us to explore the effect of such a gradient on the halo shape deduced. In this case again the best-fit is obtained for a less oblate halo of $q$ = 0.5-0.6 (compared to the $q$ = 0.4 obtained for a constant dispersion of 8 km s$^{-1}$, Section 4.4). Here ${\chi}^2$ values are lower but show a shallow minimum for the dependence on the halo shape $q$. This is because, in
this case, the observed fairly flat scale height distribution is sought to be mainly explained by a radial variation in the gas dispersion.  While this gives lower ${\chi}^2$ values as expected, it also has weak dependence on the halo shape, which is therefore not constrained well.

It is interesting that a small radial gradient in gas dispersion results in a rounder halo ($q \sim$ 0.6) which is more typical of the halo shapes seen in the cosmological simulations (Bailin \& Steinmetz 2005, Bett et al. 2007).
This shows how crucial  the value of the gas velocity is in this model, in determining the flattening of the halo, which has possible implications for the galaxy formation scenarios. This highlights the necessity of accurate measurement of gas dispersion in galaxies.

\noindent {\bf 4. Comparison with the Milky Way Galaxy:} The ratio of the total mass in the halo to the total mass in the disk (in stars and gas) within a certain radius is
an important physical quantity, since it tells one how significant the halo is at a certain radius. For the best-fit value for p=1 and q=0.4, we calculate the halo mass as a function of the radius by intergrating the halo density profile as given in equation (7). The disk mass in the two components is as observed (Section 3.2). It is now known that M31 has an extended disk upto 40 kpc with a similar disk scale length in the inner and outer regions (Ibata et al. 2005). 
We find that the fraction of the total mass in a dark matter halo is  83 \% within R= 30 kpc or about 5 R$_d$, and is 89 \% within R= 40 kpc or about 7 $R_d$.  
In comparison, for the Galaxy, using the isothermal halo model for the Galaxy and the stellar disk model as given by Mera et al. (1998) and the observed values of HI and $H_2$ gas (Scoville \& Sanders 1987), we obtain the corresponding ratio of the halo mass the total mass to be  80 \% within 5 $R_d$ (=16 kpc) , and 84 \% within 7 $R_d$ (= 22 kpc). 
These are remarkably similar in the two galaxies.

On the other hand, the scale height distribution is different in the two galaxies: it is
nearly flat increasing linearly for M31 whereas it flares in the outer Galaxy. To explain this,
 an isothermal, flattened, oblate-shaped halo is needed for M31 as shown here,
 whereas using a similar approach it was shown that 
 a spherical halo with density
falling faster than an isothermal is needed to explain the data for the Galaxy (Narayan et al. 2005). 
Thus, the dark matter halo shapes and density profiles do not appear to be universal even in large spiral galaxies.

\section{Conclusions}

We have used both the observed
 rotation curve and the outer galactic HI scale height data to constrain the dark matter halo profile of M31. We have systematically explored various shapes and power-law indices for the density distributions for the halo to fit the observed data. Our galactic disk model consists of coupled stars and HI gas, where the gas-gravity is taken into account on an equal footing with the stellar gravity. We find that an oblate isothermal halo with a central density of 0.011 M$_{\odot}$ pc$^{-3}$, a core radius of 21 kpc best fits the observations. The axis ratio for the best-fit results is 0.4.
 This is in a sharp contrast to the spherical halo used to model M31 in the literature so far. 
The rotation curve
constraint alone is usually used which determines the mass within a radius but cannot uniquely determine the shape of the halo.
The present work highlights the fact that using the two simultaneous and complementary constraints of rotation curve and the HI scale height data in the outer galactic region, allows one to identify the shape as well as the density distribution of the dark matter halo in spiral galaxies.

We stress that the availability of more data points for the HI scale heights in the outer galaxy beyond 5-6 disk scale 
lengths, and an accurate determination of the HI gas velocity dispersion, would provide a tighter  constraint for 
 the shape and the density profile of the dark matter halo. In fact, having such data for other galaxies would allow the above method to be applied to a systematic study of the dark matter halo properties in different galaxies.

\acknowledgements
We thank the anonymous referee for helpful comments which greatly
improved the presentation of results in the paper.
We thank Shashikant Gupta for his help with optimizing the numerical code.

\clearpage
\begin{figure}
\vbox to0.6in{\rule{0pt}{2.6in}}
\epsscale{.8}
\plotone{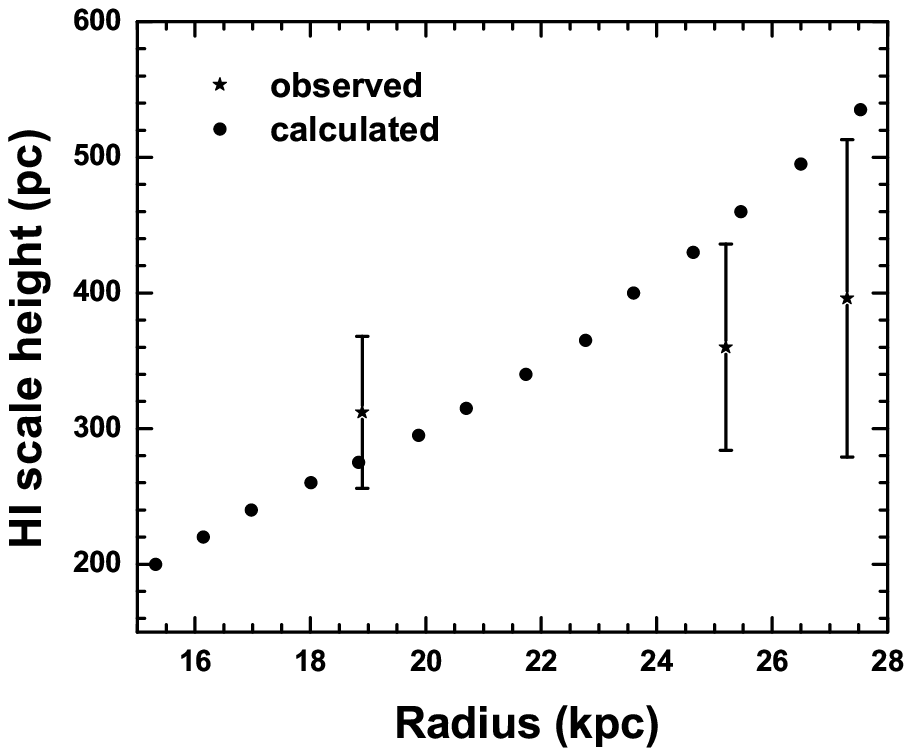}

\vskip 0.1in
\caption{Plot of the HI scale height (in pc) with the galactocentric radius (in kpc) for the best-fit case for the spherical shape ( $q$ = 1) halo for the density law $p$=1 (isothermal). Clearly, the model curve rises way above the observed distribution in the outer galaxy which is the region of interest. Thus a spherical halo is ruled out.   \label{fig1}}
\end{figure}

\clearpage
\begin{figure}
\vbox to0.6in{\rule{0pt}{2.6in}}
\epsscale{.8}
\plotone{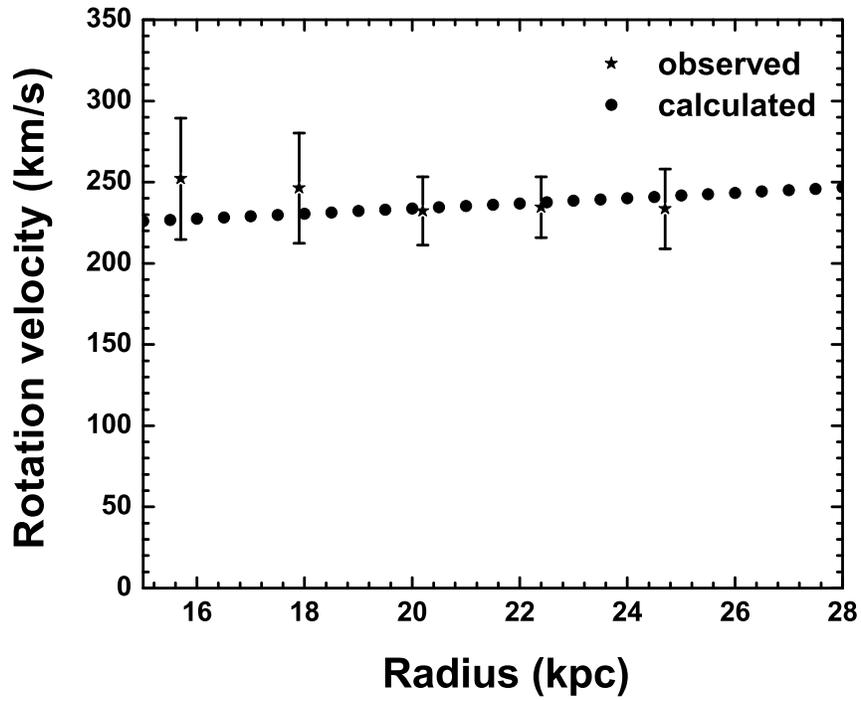}

\vskip 0.1in
\caption{Plot of the rotation velocity (in km s$^{-1}$) versus radius (in kpc) for the best-fit case of an isothermal halo of oblate shape with an axis ratio $q$ = 0.4. Our model rotation curve matches well with the observed data (Widrow et al. 2003) within the error bars. \label{fig2}}
\end{figure}

\clearpage
\begin{figure}
\vbox to0.6in{\rule{0pt}{2.6in}}
\epsscale{.8}
\plotone{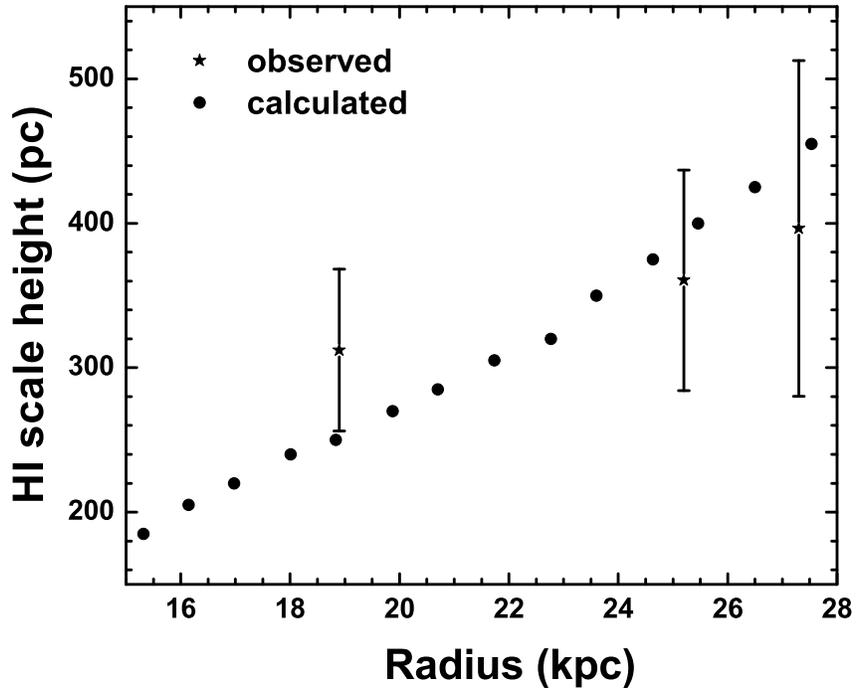}

\vskip 0.1in
\caption{ Plot of HI scale height (in pc) versus the  radius (in kpc) for the best-fit case of an isothermal halo of oblate shape with an axis ratio $q$ = 0.4. The  flattened halo of our best-fit model predicts a HI scale height distribution that agrees well with observations (Braun 1991) within the error-bars. \label{fig3}}
\end{figure}

\clearpage
\begin{figure}
\vbox to0.6in{\rule{0pt}{2.6in}}
\epsscale{.8}
\plotone{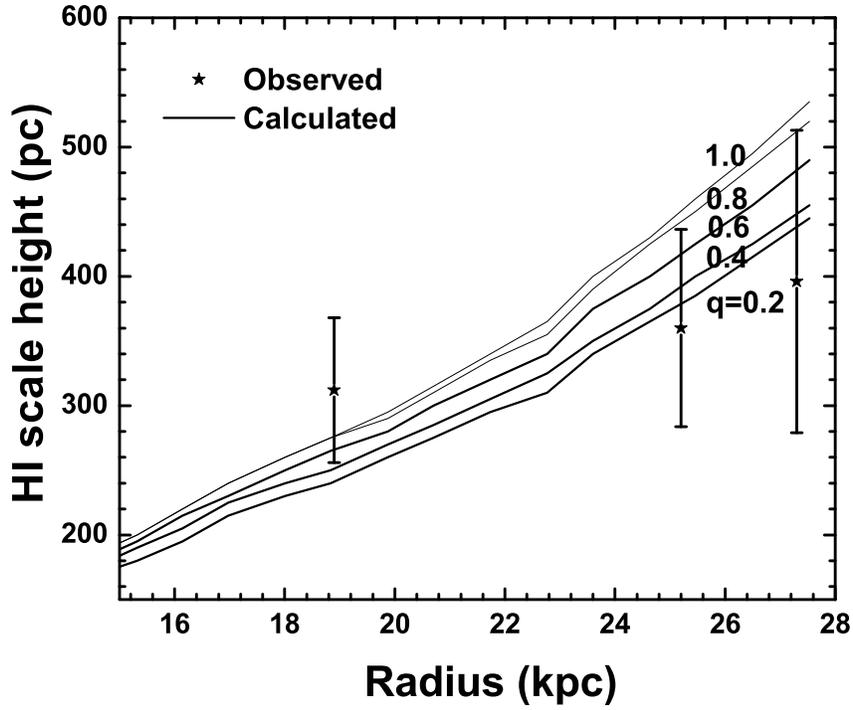}
\vskip 0.1in

\caption { Plot of the HI scale height (in pc) versus radius (in kpc) for an isothermal density profile and for different values of flattening: $q$ = 0.2, 0.4, 0.6 , 0.8 and 1, for the best-fit values of $\rho_0$ and $R_c$ in each case. This shows that a range of values between $q$ = 0.4 - 0.6 gives fits to the observed data within the error bars, but the
${\chi}^2 $ analysis identifies $q$ = 0.4 as the best-fit case.
\label{fig4}}
\end{figure}

\clearpage
\begin{figure}
\vbox to0.6in{\rule{0pt}{2.6in}}
\epsscale{.8}
\plotone{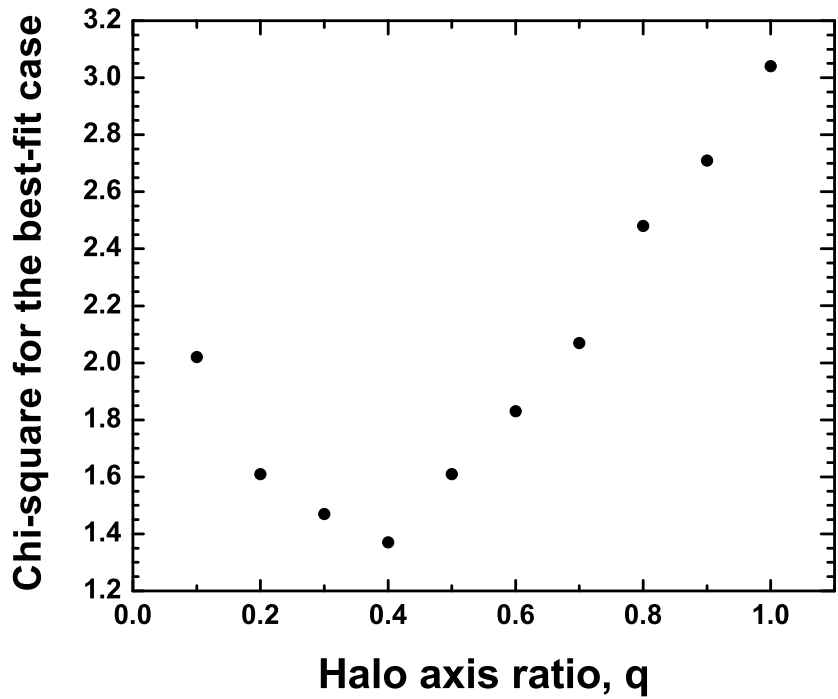}

\vskip 0.1in
\caption { Plot of ${\chi}^2$ for the best-fit case for a given axis ratio
versus the axis ratio, $q$, of the dark matter halo.
The ${\chi}^2$ shows a clear minimum at $q$ =0.4 which is thus the axis ratio that best explains the observations.
Note that   a flattened halo ($q$ = 0.4 ) is clearly distinguished from and preferred over the
spherical  case ($q$ = 1).
\label{fig5}}
\end{figure}

\clearpage
\begin{figure}
\vbox to0.6in{\rule{0pt}{2.6in}}
\epsscale{.8}
\plotone{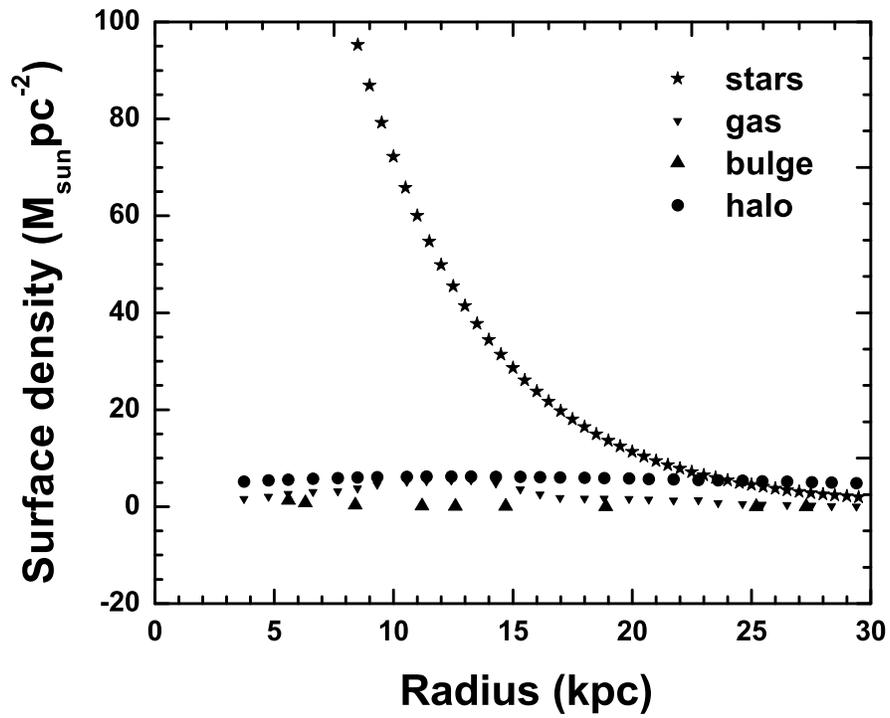}

\vskip 0.1in
\caption { Plot of stellar, gaseous, bulge and halo surface densities within the HI scale height in units of M$_{\odot}$ pc$^{-2}$ versus radius in kpc. It shows that the stellar surface density is comparable to the halo 
surface density  upto $\sim 25 $ kpc, and only beyond this the
halo starts to dominate the disk surface density as we go outward in the galaxy. 
\label{fig6}}
\end{figure}

\end{document}